\documentclass{PoS}

\usepackage{graphicx}                              % Used for includegraphics
\usepackage{amsmath,amsfonts}
\graphicspath{{./pics/}}                       % location for color fig.

\usepackage{epsfig}
\usepackage{amssymb}
\usepackage{amsfonts}
\usepackage{bbm}

\newcommand{\preprintline}{\newline
\vskip -5.2cm
\rightline{\parbox{4cm}{\large\rm  HU-EP-10/65\\ DESY 10-182}}
\vspace{4.2cm}}

\title{Effects of a potential fourth fermion generation on the Higgs boson mass bounds \preprintline}
 
\ShortTitle{Effects of a fourth fermion generation on the Higgs boson mass bounds}
 
\author{\speaker{Philipp Gerhold}\\
        Institut fur Physik, Humboldt Universit\"at zu Berlin, 12489 Berlin, Germany\\
        E-mail: \email{Philipp.Gerhold@physik.hu-berlin.de}}

\author{Karl Jansen\\
        NIC, DESY, 15738 Zeuthen, Germany\\
        E-mail: \email{Karl.Jansen@desy.de}}

\author{Jim Kallarackal\\
        Institut f\"ur Physik, Humboldt-Universit\"at zu Berlin, 12489 Berlin, Germany\\
        E-mail: \email{Jim.Kallarackal@physik.hu-berlin.de}}

\newcommand{\vs}{\vspace}
\newcommand{\hs}{\hspace}

\newcommand{\bdm}{\begin{displaymath}}
\newcommand{\edm}{\end{displaymath}}
\newcommand{\beq}{\begin{equation}}
\newcommand{\eeq}{\end{equation}}
\newcommand{\bea}{\begin{eqnarray}}
\newcommand{\eea}{\end{eqnarray}}
\newcommand{\bit}{\begin{itemize}}
\newcommand{\eit}{\end{itemize}}
\newcommand{\bc}{\begin{center}}
\newcommand{\ec}{\end{center}}
\newcommand{\re}{\relax{\rm I\kern-.18em R}}
\newcommand{\ID}{\mathbbm{1}}

\newcommand{\fhs}[1]{\mbox{\hs{#1}}}
\newcommand{\ie}{{\it i.e. }}
\newcommand{\Dov}{{\cal D}^{(ov)}}
\newcommand{\D}{\Dov}

\newcommand{\sumFL}{\sum\limits_{i=1}^{N_c}}

\newcommand{\fermiMat}{{\cal M}}

\newcommand{\latVol}[2]{\ifnum #1=#2 $#1^4$ \else $#1^3\times #2$\fi}
\newcommand{\lattice}[2]{\ifnum #1=#2 $#1^4$-lattice \else $#1^3\times #2$-lattice\fi}
\newcommand{\latticeX}[3]{\ifnum #1=#2 $#1^4$-lattice#3 \else $#1^3\times #2$-lattice#3\fi}
\newcommand{\lattices}[2]{\ifnum #1=#2 $#1^4$-lattice \else $#1^3\times #2$-lattices\fi}
\newcommand{\eq}[1]{Eq.~(\ref{#1})}
\newcommand{\eqs}[2]{Eq.~(\ref{#1}-\ref{#2})}

\newcommand{\fig}[1]{Fig.~\ref{#1}}
\newcommand{\tab}[1]{Tab.~\ref{#1}}
\newcommand{\Ref}[1]{Ref.~\cite{#1}}

\newcommand{\GEV}[1]{#1\,\mbox{GeV}}
\newcommand{\TEV}[1]{#1\,\mbox{TeV}}

\newcommand{\Nconf}{N_{Conf}}

\newcommand{\dslash}{\ensuremath\partial\kern-0.53em/}

\newcommand{\includeFigDouble}[7]{
\vs{-#6mm}
\bc
\begin{figure}[htb]
\centering
\begin{tabular}{cc}
\includegraphics[width=0.48\textwidth]{#1}
&
\includegraphics[width=0.48\textwidth]{#2}
\\
\hs{4mm}(a) & \hs{8mm}(b)  \\
\end{tabular}
\caption[#5]{#4}
\label{#3}
\vs{-2mm}
\end{figure}
\ec
\vs{-6mm}
\vs{-#7mm}
}

\newcommand{\includeTab}[5]{
\begin{table}[htb]
\centering
\begin{tabular}{#1}
\hline
#2
\hline
\end{tabular}
\caption[#5]{#4}
\label{#3}
\end{table}
}

\newcommand{\includeFigTripleDouble}[9]{
\bc
\begin{figure}[htb]
\centering
\begin{tabular}{ccc}
\includegraphics[width=0.32\textwidth]{#1}\hs{-3mm}
&
\includegraphics[width=0.32\textwidth]{#2}\hs{-3mm}
&
\includegraphics[width=0.32\textwidth]{#3}\hs{-3mm}

\\
\hs{12mm}(a) & \hs{12mm}(b) & \hs{12mm}(c) \\
\includegraphics[width=0.32\textwidth]{#4}\hs{-3mm}
&
\includegraphics[width=0.32\textwidth]{#5}\hs{-3mm}
&
\includegraphics[width=0.32\textwidth]{#6}\hs{-3mm}
\\
\hs{12mm}(d) & \hs{12mm}(e)  & \hs{12mm}(f) \\
\end{tabular}
\caption[#9]{#8}
\label{#7}
\vs{-2mm}
\end{figure}
\ec
\vs{-6mm}
}

\abstract{We study the effect of a potential fourth fermion generation on the upper and lower Higgs boson mass bounds. 
This investigation is based on the numerical evaluation of a chirally invariant lattice Higgs-Yukawa model emulating 
the same Higgs-fermion coupling structure as in the Higgs sector of the electroweak Standard Model. In particular, 
the considered model obeys a Ginsparg-Wilson version of the underlying $\mbox{SU}(2)_L\times \mbox{U}(1)_Y$ symmetry, 
being a global symmetry here due to the neglection of gauge fields in this model. We present our results 
on the modification of the upper and lower Higgs boson mass bounds induced by the presence of a hypothetical very heavy 
fourth quark doublet. Finally, we compare these findings to the standard scenario of three fermion generations.}

\FullConference{The XXVIII International Symposium on Lattice Field Theory, Lattice2010\\
		June 14-19, 2010\\
		Villasimius, Italy}

\begin{document}

\section{Introduction}
%---------------------
\label{sec:Introduction}
 
The Sakharov explanation for the matter anti-matter asymmetry of the universe suffers from the CP-violating phase 
of the Standard Model (SM3) falling short by at least 10 orders of magnitude. In addition to this concern the Sakharov 
picture demands a first order electroweak phase transition, which is also objected in the framework of the SM3. However,
both of the above caveats might be addressable~\cite{Holdom:2009rf,Carena:2004ha} by the inclusion of a new fourth 
fermion generation into an extended version of the Standard Model (SM4). Despite the arguments against the existence
of a fourth fermion generation such a scenario nevertheless remains attractive for two reasons. Firstly, there
is a strong conceptual interest, since a new fermion generation would need to be very heavy, leading to rather large 
Yukawa coupling constants and thus to potentially strong interactions with the scalar sector of the theory. Secondly,
it has been argued~\cite{Holdom:2009rf} (and the references therein) that the fourth fermion generation is actually
{\it not} excluded by electroweak precision measurements, thus leaving the potential existence of a new fermion generation
an open question.

In our contribution, however, we do not present any statement arguing in favour or disfavour of a new fermion generation. Here, 
we simply assume its existence and focus on the arising consequences on the Higgs boson mass spectrum. With the advent
of the LHC this question will become of great phenomenological interest, since the Higgs boson mass bounds, in particular
the lower bound, depend significantly on the heaviest fermion mass. Demonstrating this effect will be the main
objective of the present work. 

Due to the large Yukawa coupling constants of the fourth fermion generation a non-perturbative computation is highly 
desirable. For this purpose we employ a lattice approach to investigate the strong Higgs-fermion interaction.
In fact, we follow here the same lattice strategy that has already been used in \Ref{Gerhold:JOINT} 
for the non-perturbative determination of the upper and lower Higgs boson mass bounds in the SM3. This latter approach
has the great advantage over the preceding lattice studies of Higgs-Yukawa models that it is the first being based on 
a consistent formulation of an exact lattice chiral symmetry~\cite{Luscher:1998pq}, which allows to emulate the chiral 
character of the Higgs-fermion coupling structure of the Standard Model on the lattice
in a conceptually fully controlled manner.

\section{Numerical Results}
\label{sec:model}

In order to evaluate the Higgs boson mass bounds we have implemented a lattice 
model of the pure Higgs-fermion sector of the Standard Model. To be more precise,
the Lagrangian of the targeted Euclidean continuum model we have in mind is given as
\begin{align}
\label{eq:StandardModelYuakwaCouplingStructure}
L_{HY} &= \bar t' \dslash t' + \bar b' \dslash b' + 
\frac{1}{2}\partial_\mu\varphi^{\dagger} \partial_\mu\varphi
+ \frac{1}{2}m_0^2\varphi^{\dagger}\varphi + \lambda\left(\varphi^{\dagger}\varphi\right)^2 
+ y_{b'} \left(\bar t', \bar b' \right)_L \varphi b'_R + y_{t'} \left(\bar t', \bar b' \right)_L \tilde\varphi t'_R  \notag\\
& + \mbox{c.c. of Yukawa interactions,} 
\end{align}
where we have constrained ourselves to the consideration of the heaviest quark
doublet, \ie the fourth generation doublet, which is labeled here $(t',b')$.
This restriction is reasonable, since the dynamics of the complex scalar doublet 
$\varphi$ ($\tilde \varphi = i\tau_2\varphi^*,\, \tau_i:\, \mbox{Pauli-matrices}$)
is dominated by the coupling to the heaviest fermions. For the same reason we also 
neglect any gauge fields in this approach. The quark fields nevertheless have
a colour index which actually leads to $N_c=3$ identical copies of the fermion doublet 
appearing in the model. However, for a first exploratory study of the fermionic influence
on the Higgs boson mass bounds we have set $N_c$ to 1 for simplicity.

The actual lattice implementation of the continuum model in \eq{eq:StandardModelYuakwaCouplingStructure}
has been discussed in detail in \Ref{Gerhold:JOINT}. Since the Yukawa interaction has a 
chiral structure, it is important to establish chiral symmetry also in the lattice approach. 
This has been a long-standing obstacle, which was finally found to be circumventable by constructing 
the lattice equivalent of $\dslash$ as well as the left- and right-handed components of the quark fields 
$t'_{L,R}$, $b'_{L,R}$ on the basis of the Neuberger overlap operator~\cite{Luscher:1998pq, Neuberger:1998wv}. 
Following the proposition in \Ref{Luscher:1998pq} we have constructed a lattice Higgs-Yukawa model with a global 
$SU(2)_L \times U(1)_Y$ symmetry.

The fields considered in this model are the aforementioned doublet $\varphi$ as well as 
$N_c$ quark doublets represented by eight-component spinors $\bar\psi^{(i)}\equiv (\bar t'^{(i)}, \bar b'^{(i)})$, 
$i=1,...,N_c$. With $\D$ denoting the Neuberger overlap operator the fermionic action $S_F$ can
be written as
\bea
\label{eq:DefYukawaCouplingTerm}
S_F = \sumFL\,
\bar\psi^{(i)}\, \fermiMat\, \psi^{(i)}, 
&\quad&
\fermiMat = \D + 
P_+ \phi^\dagger \fhs{1mm}\mbox{diag}\left(y_{t'}, y_{b'}\right) \hat P_+
+ P_- \fhs{1mm}\mbox{diag}\left(y_{t'}, y_{b'}\right) \phi \hat P_-,
\eea
where $y_{t',b'}$ denote the Yukawa coupling constants and the scalar field $\varphi_x$ has been rewritten here 
as a quaternionic, $2 \times 2$ matrix $\phi^\dagger_x = (\tilde \varphi_x, \varphi_x)$, with $x$ denoting the 
site index of the $L_s^3\times L_t$-lattice. The left- and right-handed projection operators $P_{\pm}$ and the 
modified projectors $\hat P_{\pm}$ are given as
\bea
P_\pm = \frac{1 \pm \gamma_5}{2}, \quad &
\hat P_\pm = \frac{1 \pm \hat \gamma_5}{2}, \quad &
\hat\gamma_5 = \gamma_5 \left(\ID - \frac{1}{\rho} \D \right),
\eea
with $\rho$ being the radius of the circle of eigenvalues in the complex plane of the free
Neuberger overlap operator~\cite{Neuberger:1998wv}.

This action now obeys an exact global $\mbox{SU}(2)_L\times \mbox{U}(1)_Y$ 
lattice chiral symmetry. For $\Omega_L\in \mbox{SU}(2)$ and $\epsilon\in \re$ the action is invariant under the transformation
\bea
\label{eq:ChiralSymmetryTrafo1}
\psi\rightarrow  U_Y \hat P_+ \psi + U_Y\Omega_L \hat P_- \psi,
&\quad&
\bar\psi\rightarrow  \bar\psi P_+ \Omega_L^\dagger U_{Y}^\dagger + \bar\psi P_- U^\dagger_{Y}, \\
\label{eq:ChiralSymmetryTrafo2}
\phi \rightarrow  U_Y  \phi \Omega_L^\dagger,
&\quad&
\phi^\dagger \rightarrow \Omega_L \phi^\dagger U_Y^\dagger
\eea
with the compact notation $U_{Y} \equiv \exp(i\epsilon Y)$ denoting the respective representations of the 
global hypercharge symmetry group $U(1)_Y$ for the respective field it is acting on. In the continuum 
limit \eqs{eq:ChiralSymmetryTrafo1}{eq:ChiralSymmetryTrafo2} eventually recover the (here global) 
continuum $\mbox{SU}(2)_L\times \mbox{U}(1)_Y$ chiral symmetry.

Finally, the purely bosonic part $S_\varphi$ of the total lattice action $S=S_F+S_\varphi$ is given by the usual 
lattice $\varphi^4$-action
\bea
\label{eq:ContinuumPhiAction}
S_\varphi &=& \sum_{x} \frac{1}{2}\nabla^f_\mu\varphi_x^{\dagger} \nabla^f_\mu\varphi_x
+ \frac{1}{2}m_0^2\varphi_x^{\dagger}\varphi_x + \lambda\left(\varphi_x^{\dagger}\varphi_x\right)^2 ,
\eea
with the bare mass $m_0$, the forward difference operator $\nabla^f_\mu$ in direction $\mu$, and the bare quartic 
coupling constant $\lambda$. For the practical lattice implementation, however, a reformulation of \eq{eq:ContinuumPhiAction}
in terms of the hopping parameter $\kappa$ and the lattice quartic coupling constant $\hat\lambda$ proves to be more
convenient. It reads
\beq
\label{eq:LatticePhiAction}
S_\Phi = -\kappa\sum_{x,\mu} \Phi_x^{\dagger} \left[\Phi_{x+\mu} + \Phi_{x-\mu}\right]
+ \sum_{x} \Phi^{\dagger}_x\Phi_x + \hat\lambda \sum_{x} \left(\Phi^{\dagger}_x\Phi_x - N_c \right)^2,
\eeq
and is fully equivalent to \eq{eq:ContinuumPhiAction}. This alternative formulation opens the possibility of explicitly
studying the limit $\lambda = \infty$ on the lattice. The aforementioned connection can be established through a rescaling 
of the scalar field $\Phi_x \in \re^4$ and the involved coupling constants according to
\beq
\label{eq:RelationBetweenHiggsActions}
\varphi_x = \sqrt{2\kappa}
\left(
\begin{array}{*{1}{c}}
\Phi_x^2 + i\Phi_x^1\\
\Phi_x^0-i\Phi_x^3\\ 
\end{array}
\right), 
\quad
\lambda=\frac{\hat\lambda}{4\kappa^2}, \quad
m_0^2 = \frac{1 - 2N_c\hat\lambda-8\kappa}{\kappa}.
\eeq

Due to the triviality of the Higgs sector the targeted Higgs boson mass bounds actually depend on the 
non-removable, intrinsic cutoff parameter $\Lambda$ of the considered Higgs-Yukawa theory, which can be
defined as the inverse lattice spacing, \ie $\Lambda=1/a$. To determine these cutoff dependent bounds
for a given phenomenological scenario, \ie for given hypothetical masses of the fourth fermion generation,
the strategy is to evaluate the maximal interval of Higgs boson masses attainable within the framework of 
the considered Higgs-Yukawa model being in consistency with this phenomenological setup. The free parameters 
of the model, being the bare scalar mass $m_0$, the bare quartic coupling constant $\lambda$, and the Yukawa 
coupling constants $y_{t',b'}$ thus have to be tuned accordingly. The idea is to use the phenomenological
knowledge of the renormalized vacuum expectation value $v_r/a = \GEV{246}$ of the scalar field $\varphi$ as 
well as the hypothetical fourth generation quark masses $m_{t',b'}$ to fix the bare model parameters for a 
given cutoff $\Lambda$.

In lack of an additional matching condition a one-dimensional freedom is left open here, which can be parametrized
in terms of the quartic coupling constant $\lambda$. This freedom finally leads to the emergence of upper and lower
bounds on the Higgs boson mass. As expected from perturbation theory, one also finds numerically~\cite{Gerhold:JOINT} 
that the lightest and heaviest Higgs boson masses are obtained at vanishing and infinite bare quartic coupling constant, 
respectively. The lower mass bound will therefore be obtained at $\lambda=0$, while $\lambda=\infty$ will be adjusted to 
derive the upper bound.

Concerning the hypothetical masses of the fourth fermion generation quarks, we target here a mass degenerate scenario 
with $m_{t'}/a=m_{b'}/a=\GEV{700}$, which is somewhat above its tree-level unitarity upper bound~\cite{Chanowitz:1978mv}.
However, we are currently also investigating a set of other mass settings to study in particular the quark mass dependence of
the Higgs boson mass bounds.

\includeTab{|ccccccc|}
{
$\kappa$ & $L_s$       & $L_t$ &  $N_c$ &  $m_0^2$             & $\lambda$      & $y_{t'}=y_{b'}$      \\
\hline
0.09442  & 12,16,20,24 &   32  &  1     &  2.5910               & 0              & 3.2122                \\
0.09485  & 12,16,20,24 &   32  &  1     &  2.5430               & 0              & 3.2049                \\
0.09545  & 12,16,20,24 &   32  &  1     &  2.4767               & 0              & 3.1949                \\
0.09560  & 12,16,20,24 &   32  &  1     &  2.4603               & 0              & 3.1923                \\
0.09605  & 12,16,20,24 &   32  &  1     &  2.4112               & 0              & 3.1849                \\ \hline
0.21300  & 12,16,20,24 &   32  &  1     &  $\infty$            & $\infty$        & 3.3707                \\
0.21500  & 12,16,20,24 &   32  &  1     &  $\infty$            & $\infty$        & 3.3550                \\
0.22200  & 12,16,20,24 &   32  &  1     &  $\infty$            & $\infty$        & 3.1816                \\
0.22320  & 12,16,20,24 &   32  &  1     &  $\infty$            & $\infty$        & 3.1730                \\
0.22560  & 12,16,20,24 &   32  &  1     &  $\infty$            & $\infty$        & 3.1561                \\
}
{tab:SummaryOfParametersForUpperHiggsMassBoundRuns}
{The model parameters underlying the lattice calculations performed in this study are presented. The setting $\lambda=0$ ($\lambda=\infty$) is employed
for deriving the lower (upper) Higgs boson mass bound. Depending on the lattice volume the available statistics ranges from $\Nconf=1,000$ to
$\Nconf=20,000$.
}
{Model parameters of the Monte-Carlo runs underlying the lattice calculation of the upper and lower Higgs boson mass bounds.}

For the eventual determination of the cutoff dependent Higgs boson mass bounds several series of Monte-Carlo calculations 
have been performed at different values of $\Lambda$ and on different lattice volumes as summarized in \tab{tab:SummaryOfParametersForUpperHiggsMassBoundRuns}.
In order to tame finite volume effects as well as cutoff effects to an acceptable level, we have demanded as 
a minimal requirement that all particle masses $\hat m=m_{H}, m_{t'}, m_{b'}$ in lattice units fulfill $\hat m < 0.5$ 
and $\hat m\cdot L_{s,t}>3.5$, at least on the largest investigated lattice volumes. Assuming the Higgs boson mass $m_H$ 
to be around $\GEV{500-750}$ this allows to reach cutoff scales between $\GEV{1500}$ and $\GEV{3500}$ on a \latticeX{24}{32}{.} 
However, despite this setting strong finite volume effects are nevertheless expected induced by the massless Goldstone modes. 
It is known that these finite size effects are proportional to $1/L_s^2$ at leading order. An infinite volume extrapolation 
of the lattice data is therefore mandatory. 

\includeFigTripleDouble{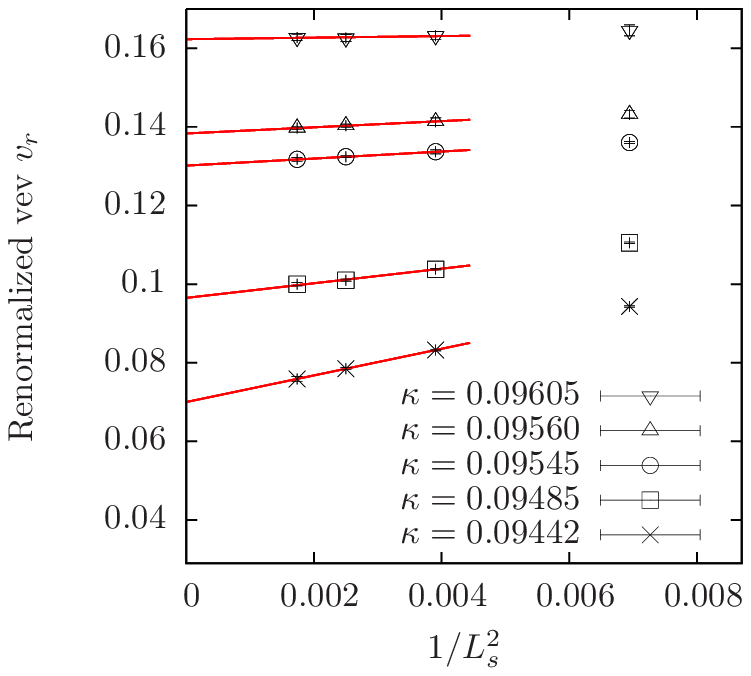}{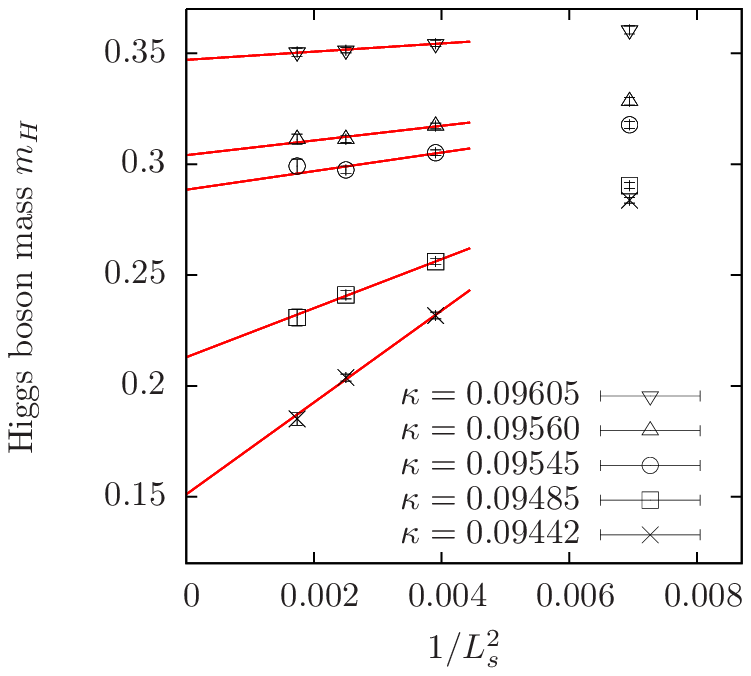}{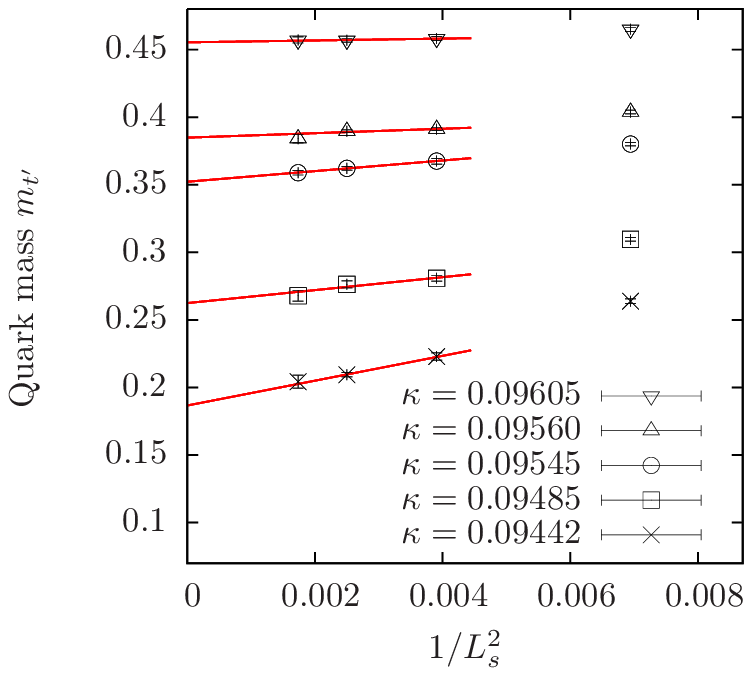}
{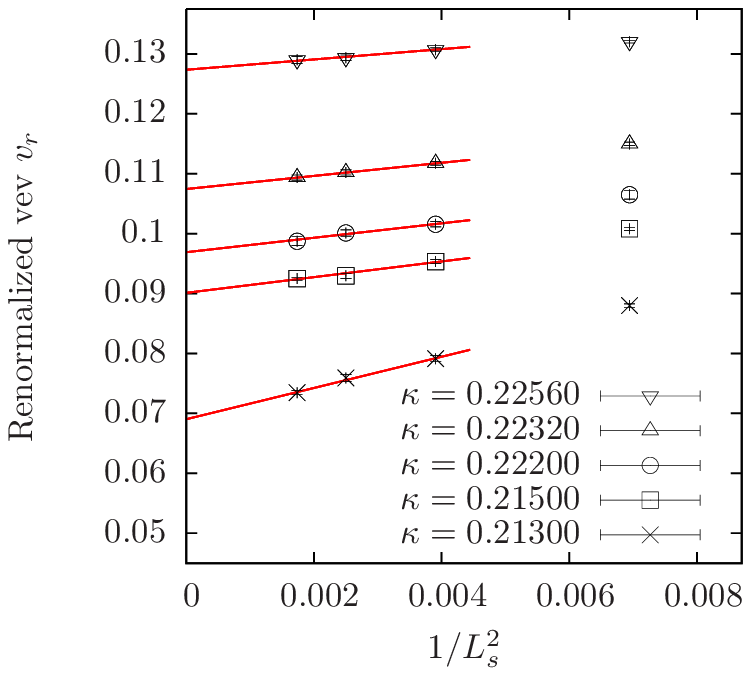}{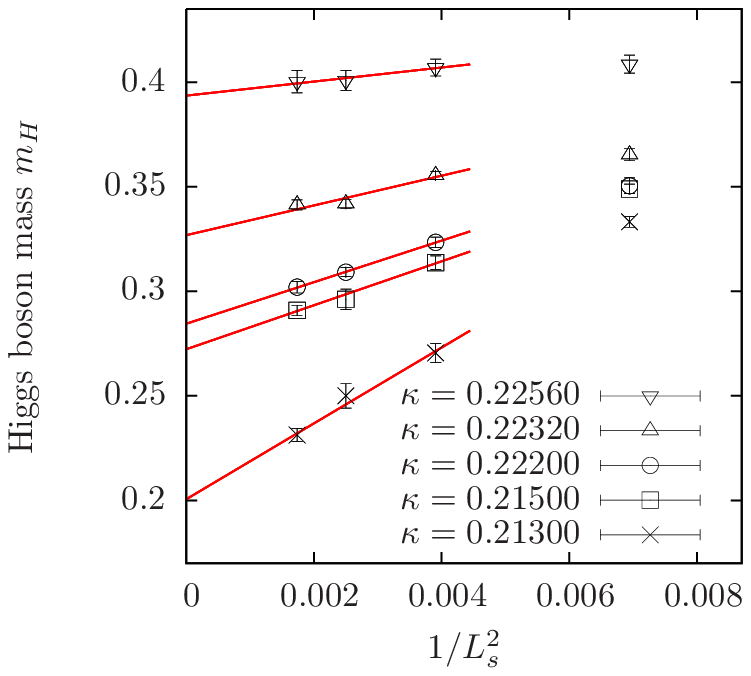}{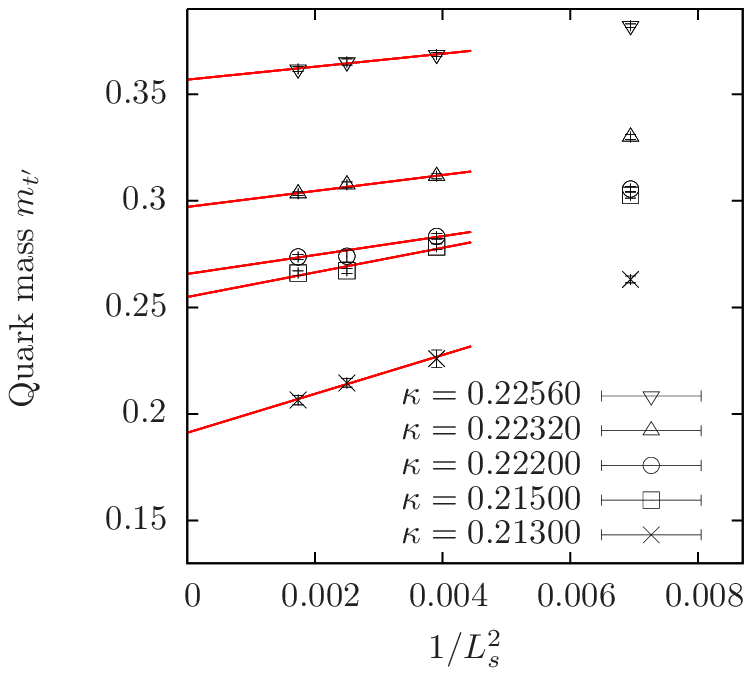}
{fig:FiniteSizeEffects}
{The finite volume data of the renormalized vacuum expectation value $v_r$ (a,d), the Higgs boson mass $m_H$ (b,e), and the degenerate quark mass 
$m_{t'}=m_{b'}$ (c,f), as obtained from the lattice calculations specified in \tab{tab:SummaryOfParametersForUpperHiggsMassBoundRuns},
are plotted versus $1/L_s^2$. The upper (lower) row corresponds to the setting $\lambda=0$ ($\lambda=\infty$). The infinite volume 
extrapolation is performed by fitting the data to a linear function. Due to the observed curvature arising from the non-leading finite
volume corrections only those data with $L_s\ge 16$ have been respected by the linear fit procedures.
}
{FiniteVolumeLatticeDataLowerBound}

The finite volume data of the renormalized vacuum expectation value $v_r$, the Higgs boson mass $m_H$, and the degenerate quark mass 
$m_{t'}=m_{b'}$ resulting from the calculations specified in \tab{tab:SummaryOfParametersForUpperHiggsMassBoundRuns}
are presented in \fig{fig:FiniteSizeEffects}. For the details about the determination of the latter observables the interested
reader is referred to \Ref{Gerhold:JOINT}. Here it is only stated that the renormalization constant $Z_G$ entering the
renormalization of the scalar field $\varphi$ has been derived from the Goldstone propagator, the Higgs boson mass $m_H$
has been taken from the Higgs propagator, and the quark masses have been computed from the time correlation function of
the respective fermionic fields. These lattice data are plotted versus $1/L_s^2$ and extrapolated to the infinite volume limit 
by means of a linear fit ansatz according to the aforementioned leading order behaviour. Due to the observed curvature arising 
from the non-leading finite volume corrections only those data with $L_s\ge 16$ have been respected by the linear fit procedures. 
One finds that the intended infinite volume extrapolation can indeed reliably be performed thanks to the multitude of investigated 
lattice volumes reaching from $12^3\times 32$ to \lattices{24}{32} here.

The quality of the tuning procedure intending to hold the quark masses constant can then be examined in \fig{fig:PhysicalTopMasses}b displaying
the results of the infinite volume extrapolation of $m_{t'}$. In the considered SM4 scenario the fluctuation of the quark mass has been constrained 
to roughly $m_{t'}=m_{b'}=\GEV{676\pm 22}$. For later comparisons with the corresponding SM3 scenario we also present the analogous summary plot of 
our earlier investigations~\cite{Gerhold:JOINT} of the latter setup where the degenerate quark masses have been fixed to approximately 
$m_t=m_b=\GEV{173\pm 3}$ as demonstrated in \fig{fig:PhysicalTopMasses}a.

\includeFigDouble{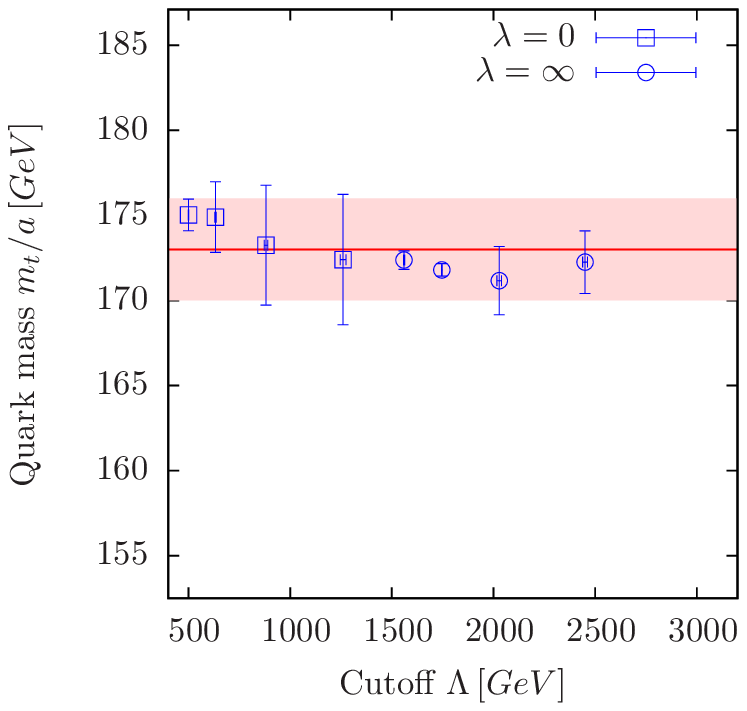}{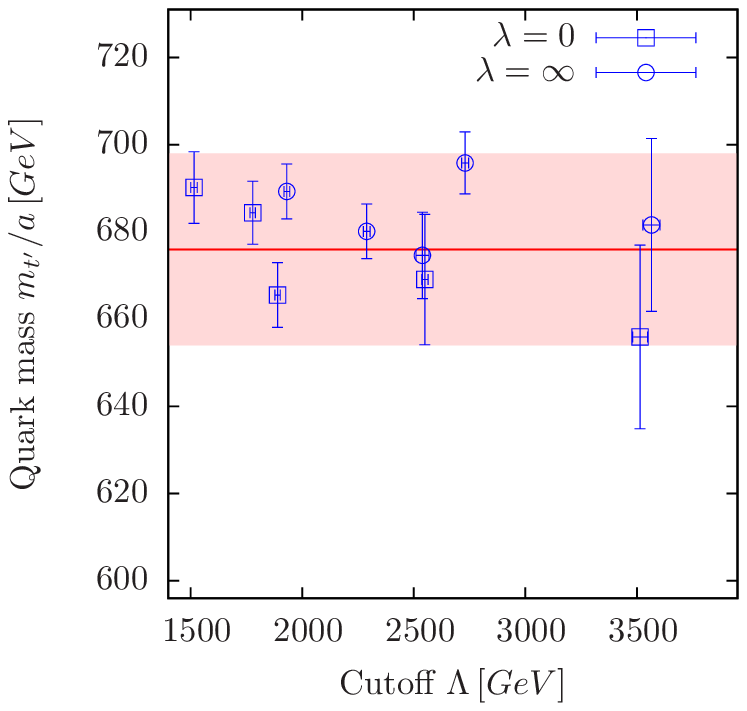}
{fig:PhysicalTopMasses}
{The infinite volume extrapolations of the degenerate quark masses observed in the lattice calculations specified in \tab{tab:SummaryOfParametersForUpperHiggsMassBoundRuns}
are presented versus the cutoff parameter $\Lambda$. In the SM3 scenario (a) the fluctuation of the quark mass has been constrained to $m_t=m_b=\GEV{173\pm 3}$, 
while $m_{t'}=m_{b'}=\GEV{676\pm 22}$ is adjusted in the SM4 scenario (b). 
}
{InfiniteVolumeExtrapolationBothBounds}{3}{1} 

The infinite volume results of the Higgs boson masses are finally presented in \fig{fig:PhysicalHiggsMassBounds}b. The numerical data for the
upper mass bound have moreover been fitted with the analytically expected functional form of the cutoff dependence of the upper
Higgs boson mass bound derived in \Ref{Luscher:1988uq}. It is given as
\bea
\label{eq:StrongCouplingLambdaScalingBeaviourMass}
\frac{m^{up}_{H}}{a} &=& A_m \cdot \left[\log(\Lambda^2/\mu^2) + B_m \right]^{-1/2}, 
\eea
with $A_m$, $B_m$ denoting the free fit parameters and $\mu$ being an arbitrary scale, which we have chosen as $\mu=\TEV{1}$ here. One 
learns from this presentation that the obtained results are indeed in good agreement with the expected logarithmic decline of the upper Higgs 
boson mass bound with increasing cutoff parameter $\Lambda$.

The reader may want to compare these findings to the upper and lower Higgs boson mass bounds previously derived in the SM3. The lattice 
results corresponding to that setup have been determined in \Ref{Gerhold:JOINT} and are summarized in \fig{fig:PhysicalHiggsMassBounds}a.
The main finding is that especially the lower bound is drastically shifted towards larger values in the presence of the assumed mass-degenerate 
fourth quark doublet. From this analysis it can be concluded that the usually expected light Higgs boson seems to be 
incompatible with a very heavy fourth fermion generation.

\includeFigDouble{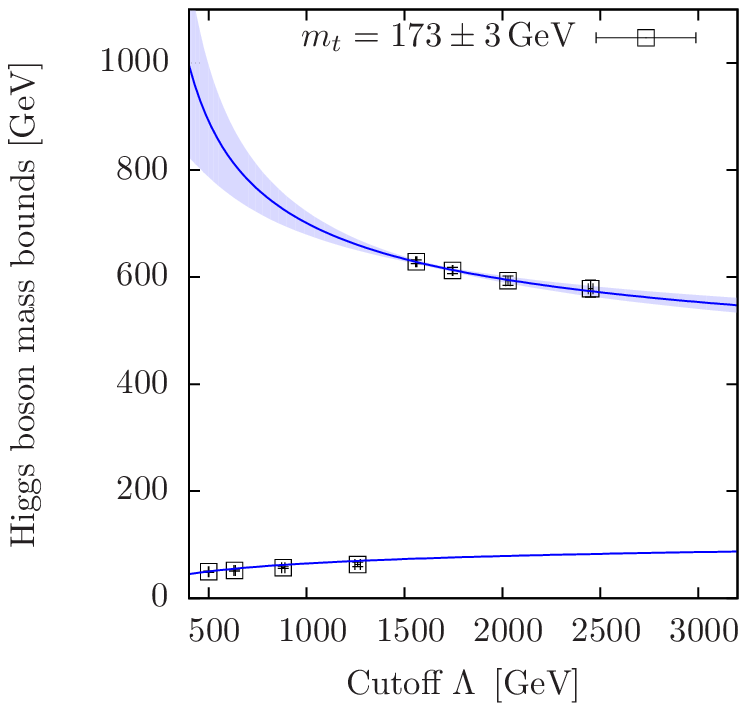}{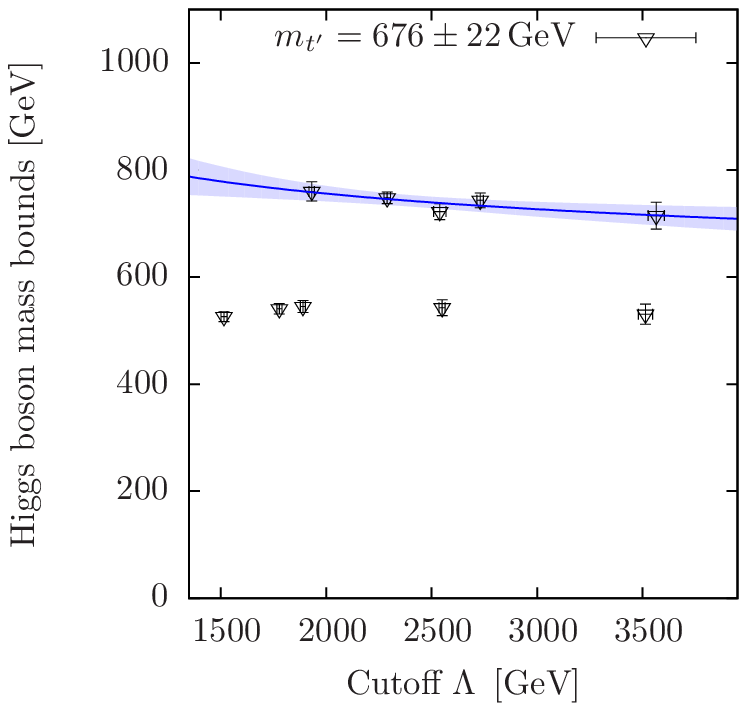}
{fig:PhysicalHiggsMassBounds}
{Upper and lower Higgs boson mass bounds are shown for $N_c=1$, $m_t=m_b=\GEV{173\pm 3}$ (a) and $N_c=1$, $m_{t'}=m_{b'}=\GEV{676\pm 22}$ (b).
Both upper bounds are each fitted with \eq{eq:StrongCouplingLambdaScalingBeaviourMass}. The lower bound in (a)
is also compared to a direct analytical computation depicted by the solid line as discussed in \Ref{Gerhold:JOINT}.
}
{InfiniteVolumeExtrapolationBothBounds}{3}{1}

\section*{Acknowledgments}
We thank George Hou and David Lin for ongoing discussions and M. M\"uller-Preussker for his continuous support.
We moreover acknowledge the support of the DFG through the DFG-project {\it Mu932/4-2}.
The numerical computations have been performed on the {\it HP XC4000 System}
at the {Scientific Supercomputing Center Karlsruhe} and on the
{\it SGI system HLRN-II} at the {HLRN Supercomputing Service Berlin-Hannover}.

\bibliographystyle{unsrtOwnNoTitles}
\bibliography{Proceedings}

\begin{thebibliography}{1}

\bibitem{Holdom:2009rf}
B.~Holdom, W.~S. Hou, T.~Hurth, M.~L. Mangano, S.~Sultansoy, et~al.
\newblock {\em PMC Phys.}, A3:4, 2009.

\bibitem{Carena:2004ha}
M.~S. Carena, A.~Megevand, M.~Quiros, and C.~E.~M. Wagner.
\newblock {\em Nucl.Phys.}, B716:319--351, 2005.

\bibitem{Gerhold:JOINT}
P.~Gerhold and K.~Jansen.
\newblock {{\it JHEP}, 0710:001, 2007. {\it JHEP}, 0709:041, 2007. {\it JHEP},
  0907:025, 2009. {\it JHEP}, 1004:094, 2010. {\it arXiv}, 1002.2569, 2010}.

\bibitem{Luscher:1998pq}
M.~L{\"u}scher.
\newblock {\em Phys. Lett.}, B428:342--345, 1998.

\bibitem{Neuberger:1998wv}
H.~Neuberger.
\newblock {\em Phys. Lett.}, B427:353--355, 1998.

\bibitem{Chanowitz:1978mv}
M.~S. Chanowitz, M.~A. Furman, and I.~Hinchliffe.
\newblock {\em Nucl. Phys.}, B153:402, 1979.

\bibitem{Luscher:1988uq}
M.~L{\"u}scher and P.~Weisz.
\newblock {\em Nucl. Phys.}, B318:705, 1989.

\end{thebibliography}

\end{document}